Relativistic five-quark equations and negative parity pentaquarks.

Gerasyuta S.M.[1,2], Kochkin V.I.[1]

1. Department of Theoretical Physics, St. Petersburg State University, 198904, St. Petersburg, Russia.
2. Department of Physics, LTA, 194021, St. Petersburg, Russia,
   E-mail: gerasyuta@sg6488.spb.edu

Abstract

The relativistic five-quark equations are found in the framework of the dispersion relation technique. The solutions of these equations using the method based on the extraction of the leading singularities of the amplitudes are obtained. The five-quark amplitudes for the low-lying pentaquarks including the u, d, s- quarks are calculated. The poles of these amplitudes determine the masses of the negative parity pentaquarks with I = 0, 1 and $J^P = \frac{3}{2}^-, \frac{5}{2}^-$. The mass of the lowest pentaquark with I = 0 and $J^P = \frac{3}{2}^-$ is equal to 1514 MeV.

I. Introduction.

Recent observation of an exotic baryon state with positive strangeness, $\theta^+$(1540), by LEPS collaboration in Spring – 8 [1] and subsequent experiments [2 - 10] has raised great interest in hadron physics. This state cannot be an ordinary three-quark baryon since it has positive strangeness, and therefore the minimal quark content is $udud\bar{s}$. The most striking feature of $\theta^+$(1540) is that the width is unusually small: $\Gamma <$ 25 MeV despite the fact that it lies about 100 MeV above the $NK$ threshold.

The discovery of $\theta^+$ has triggered intensive theoretical studies to understand the structure of the $\theta^+$ [11 - 20]. One of the main issues is to clarify the quantum numbers, especially, the spin and the parity, which are key properties to understand the abnormally small width. Recent analysis of the $K^+$ scattering from the xenon or deuteron implies even smaller value $\Gamma <$ 1 MeV [21 - 23]. The chiral soliton model has predicted the masses and widths of the pentaquark baryons with less theoretical ambiguity based on the $SU(3)_f$ flavor algebra [24]. The model indicates the width of $\theta^+$ around a few ten MeV [25]. In the non-relativistic quark model (A. Hosaka et al) [26] found that the negative parity $\theta^+$ width becomes very large which is of order of several hundreds MeV, while it is about a several tens MeV for the positive parity. By assuming additionally diquark correlations, the width is reduced to be of order 10 MeV.

However the possibility of $J^P = \frac{3}{2}^-$ is not excluded. If $\theta^+$ carries $J^P = \frac{3}{2}^-$, it decays into nucleons and kaons in the D-wave. The phase space suppression factor is $\sim (p_k / m_\theta)^5 \sim 10^{-5}$, where $p_k$ is the decay momentum of kaon. The decay width could be well below ten MeV even if the coupling constant $g_{\theta KN}$ is big due to $\theta$ s strong overlap with $KN$.

There have been a few theoretical papers on the possible $J^P = \frac{3}{2}^-$ pentaquarks using different models. Page et al. suggested I = 2, $J^P = \frac{1}{2}^-, \frac{3}{2}^-$ for $\theta^+$ to resolve the narrow width puzzle [17]. Jaffe and Wilczek discussed the $J^P = \frac{3}{2}^-$ assignment for $\Xi$ pentaquark [27]. The mass spectrum of $J^P = \frac{3}{2}^-$ pentaquarks were studied with the perturbative chiral quark model [28]. Takeuchi and Shimizu suggested the observed $\theta$ resonance as I = 0, $J^P = \frac{3}{2}^-$ $NK^*$ bound state using the quark model [29]. With the flux tube model, Kanada-Enyo et al. studied the mass and decay width of the I = 1, $J^P = \frac{3}{2}^-$ pentaquark [30]. Huang et al. proposed $\theta^+$ as molecular state of $NK\pi$ with I = 1, $J^P = \frac{3}{2}^-$ using the chiral $SU(3)$ quark model [31]. The phenomenology of $J^P = \frac{3}{2}$ pentaquarks such as the mixing scheme and mass pattern was discussed in [32]. Pentaquark states with $J^P = \frac{3}{2}$ and I = 0, 1 were studied using currents composed of one scalar diquark and one vector diquark [33]. Shi-Lin Zhu investigated the possible existence of the spin 3/2 pentaquark states using the finite energy sum rule [34].

In our previous paper the relativistic five-quark equations for the family of the $\theta$ pentaquarks are constructed [35]. The five-quark amplitudes for the low-lying $\theta$ pentaquarks are calculated. The poles of these amplitudes determine the masses of the $\theta$ pentaquarks. The masses of the constituent u, d, s -quarks coincide with the quark masses of the ordinary baryons in our quark model [36]: $m_{u,d}$ = 410 MeV, $m_s$ = 557 MeV. The model has only three parameters. The cut-off parameters $\Lambda_{0^+}$ =16.5 and $\Lambda_{1^+}$ =20.12 are similar to those [37]. The gluon coupling constant $g$ =0.456 is fitted by fixing $\theta^+$ pentaquark mass $m$(1540). We received the isotensor $\theta$ pentaquarks with $J^P = \frac{1}{2}^\pm, \frac{3}{2}^\pm$ and predict the masses of $\theta^{++}$ ($\theta^0$) and $\theta^{+++}$ ($\theta^-$) pentaquarks [35].

The present paper is devoted to the construction of relativistic five-quark equations for the low-lying negative parity pentaquarks, which decay into nucleons and kaons in the D-wave. The pentaquarks built with $\bar{s}$ quark in D-wave. The five-quark amplitudes for the negative parity pentaquarks are constructed. The poles of these amplitudes determine the masses of the negative parity pentaquarks with I = 0, 1 and $J^P = \frac{3}{2}^-, \frac{5}{2}^-$. The mass of the lowest pentaquark with I = 0 and $J^P = \frac{3}{2}^-$ is equal to 1514 MeV. We use the parameters of the previous paper.

The paper is organized as follows. After this introduction, we discuss the five-quark amplitudes, which contain $\bar{s}$ - antiquark in the D-wave and four nonstrange quarks (Section 2). In section 3, we report our numerical results (Tables I, II) and the last section is devoted to the discussion and conclusion.

II. Five-quark amplitudes for the negative parity pentaquarks.

We derive the relativistic five-quark equations in the framework of the dispersion relation technique. We use only planar diagrams; the other diagrams due to the rules of $1/N_c$ expansion [38 - 40] are neglected.

The current generates a five-quark system. Their successive pair interactions lead to the diagrams shown in Figs. 1, 2. The correct equations for the amplitude are obtained by taking into account all possible subamplitudes. It corresponds to the division of complete system into subsystems smaller number of particles. Then one should represent a five-particle amplitude as a sum of ten subamplitudes:

$$A = A_{12} + A_{13} + A_{14} + A_{15} + A_{23} + A_{24} + A_{25} + A_{34} + A_{35} + A_{45}. \qquad (1)$$

This defines the division of the diagrams into group according to the certain pair interaction of particles. The total amplitude can be represented graphically as a sum of diagrams.

We need to consider only one group of diagrams and the amplitude corresponding to them, for example $A_{12}$. We shall consider the derivation of the relativistic generalization of the Faddeev-Yakubovsky approach for $\theta^+$ pentaquark. We shall construct the five-quark amplitude of four up quarks and one strange antiquark, in which the pair interactions with the quantum numbers of $0^+$ and $1^+$ diquarks are included. The set of diagrams associated with the amplitude $A_{12}$ can further be broken down into groups corresponding to amplitudes: $A_1(s, s_{1234}, s_{12}, s_{34})$, $A_2(s, s_{1234}, s_{12}, s_{35})$, $A_3(s, s_{1234}, s_{25}, s_{34})$, $A_4(s, s_{1234}, s_{24}, s_{35})$, $A_5(s, s_{1234}, s_{13}, s_{134})$, $A_6(s, s_{1234}, s_{23}, s_{234})$, $A_7(s, s_{1234}, s_{24}, s_{234})$ (Fig. 1). The $\bar{s}$ is shown by the arrow and other lines correspond to the four nonstrange quarks. In the case of the $\theta^{++}$ ($\theta^0$) pentaquarks we need to use six (Fig. 2) subamplitudes.

Here $s_{ik}$ is the two-particle subenergy squared, $s_{ijk}$ corresponds to the energy squared of particles $i$, $j$, $k$, $s_{ijkl}$ is the four-particle subenergy squared and $s$ is the system total energy squared.

The system of graphical equations are determined by the subamplitudes using the self-consistent method. If we consider the $\theta^+$ state, we use seven subamplitudes (Fig. 1). In the case of the $\theta^{++}$ ($\theta^0$) pentaquarks we derive six (Fig. 2) subamplitudes. The coefficients are determined by the permutation of quarks [41, 42]. We do not extract the initial four- and three-particle singularities in the subamplitude $A_1(s, s_{1234}, s_{12}, s_{34})$ because they are weaker than the two-particle singularities.

In order to represent the subamplitudes $A_1(s, s_{1234}, s_{12}, s_{34})$, $A_2(s, s_{1234}, s_{12}, s_{35})$, $A_3(s, s_{1234}, s_{25}, s_{34})$, $A_4(s, s_{1234}, s_{24}, s_{35})$, $A_5(s, s_{1234}, s_{13}, s_{134})$, $A_6(s, s_{1234}, s_{23}, s_{234})$, $A_7(s, s_{1234}, s_{24}, s_{234})$ in the form of a dispersion relation it is necessary to define the amplitudes of quark-quark and quark-antiquark interaction $b_n(s_{ik})$. The pair quarks amplitudes $q\bar{q} \to q\bar{q}$ and $qq \to qq$ are calculated in the framework of the dispersion N/D method with the input four-fermion interaction [43- 45] with quantum numbers of the gluon [46]. The regularization of the dispersion integral for the D-function is carried out with the cut-off parameters $\Lambda_n$.

The four-quark interaction are considered as an input [46]:

$$g_v(\bar{q}\vec{\lambda} I_f \gamma_\mu q)^2 + 2g_v^{(s)}(\bar{q}\lambda\gamma_\mu I_f q)(\bar{s}\lambda\gamma_\mu s) + g_v^{(ss)}(\bar{s}\lambda\gamma_\mu s)^2 \tag{2}$$

Here $I_f$ is the unity matrix in the flavor space (u, d), $\lambda$ are the color Gell-Mann matrices. Dimentional constants of the four-fermion interaction $g_v$, $g_v^{(s)}$ and $g_v^{(ss)}$ are parameters of the model. At $g_v = g_v^{(s)} = g_v^{(ss)}$ the flavor $SU(3)_f$ symmetry occurs. The strange quark violates the flavor $SU(3)_f$ symmetry. In order to avoid an additional violation parameters we introduce the scale shift of the dimentional parameters [46]:

$$\gamma = \frac{m^2}{\pi^2} g_v = \frac{(m+m_s)^2}{4\pi^2} g_v^{(s)} = \frac{m_s^2}{\pi^2} g_v^{(ss)} \tag{3}$$

$$\lambda_n = \frac{4\Lambda_n(ik)}{(m_i + m_k)^2}, \text{ n}= 1, 2, 3. \tag{4}$$

Here $m_i$ and $m_k$ are the quark masses in the intermediate state of the quark loop. Dimensionless parameters $\gamma$ and $\lambda_n$ are supposed to be constants which are independent of the quark interaction type. The applicability of (Eq. 2) is verified by the success of De Rujula-Georgi-Glashow quark model [47], where only the short-range part of Breit potential connected with the gluon exchange is responsible for the mass splitting in hadron multiplets.

We use the results of our relativistic quark model [46] and write down the pair quarks amplitude in the form:

$$b_n(s_{ik}) = \frac{G_n^2(s_{ik})}{1 - B_n(s_{ik})} \qquad (5)$$

$$B_n(s_{ik}) = \int_{(m_1+m_2)^2}^{\Lambda_n} \frac{ds'_{ik}}{\pi} \frac{\rho_n(s'_{ik}) G_n^2(s'_{ik})}{s'_{ik} - s_{ik}}. \qquad (6)$$

Here $G_n(s_{ik})$ are the quark-quark and quark-antiquark vertex functions (Table III). The vertex functions are determined by the contribution of the crossing channels. These vertex functions satisfy the Fierz relations. The all of these vertex functions are generated from $g_V$, $g_V^{(s)}$ and $g_V^{(ss)}$. $B_n(s_{ik})$, $\rho_n(s_{ik})$ are the Chew-Mandelstam function with cut-off $\Lambda_n$ ($\Lambda_1 = \Lambda_3$) [48] and the phase space respectively:

$$\rho_n(s_{ik}, J^{PC}) = \left( \alpha(J^{PC}, n) \frac{s_{ik}}{(m_i + m_k)^2} + \beta(J^{PC}, n) \right) \frac{\sqrt{[s_{ik} - (m_i + m_k)^2][s_{ik} - (m_i - m_k)^2]}}{s_{ik}},$$

The coefficients $\alpha(J^{PC}, n)$ and $\beta(J^{PC}, n)$ are given in Table IV. Here n=1 corresponds to a $qq$-pair with $J^P = 0^+$ in the $\bar{3}_c$ color state, n=2 describes a $qq$-pair with $J^P = 1^+$ in the $\bar{3}_c$ color state and n=3 defines the $q\bar{q}$-pairs corresponding to meson with quantum numbers: $J^P = 2^-$.

In the case in question the interacting quarks do not produce a bound state, therefore the integration in Eqs. (7) - (13) is carried out from the threshold $(m_i + m_k)^2$ to the cut-off $\Lambda_n$. The system of integral equations systems, corresponding to Fig. 1 (the meson state with $J^P = 2^-$ and diquarks with $J^P = 0^+, 1^+$) can be described as:

$$A_1(s, s_{1234}, s_{12}, s_{34}) = \frac{\lambda_1 B_3(s_{12}) B_1(s_{34})}{[1 - B_3(s_{12})][1 - B_1(s_{34})]} + 3\hat{J}_2(3,1) A_6(s, s_{1234}, s'_{23}, s'_{234}) +$$
$$+ 3\hat{J}_2(3,1) A_7(s, s_{1234}, s'_{24}, s'_{234}) + 2\hat{J}_2(3,1) A_5(s, s_{1234}, s'_{13}, s'_{134}) + 2\hat{J}_1(3) A_5(s, s_{1234}, s'_{15}, s_{125}) +, \quad (7)$$
$$+ 2\hat{J}_1(3) A_6(s, s_{1234}, s'_{25}, s_{125}) + 2\hat{J}_1(1) A_7(s, s_{1234}, s'_{34}, s_{345}) + 2\hat{J}_1(1) A_6(s, s_{1234}, s'_{34}, s_{345})$$

$$A_2(s,s_{1234},s_{12},s_{35}) = \frac{\lambda_2 B_3(s_{12}) B_2(s_{35})}{[1-B_3(s_{12})][1-B_2(s_{35})]} + 6\hat{J}_2(3,2) A_6(s,s_{1234},s'_{23},s'_{234}) +$$
$$+ 2\hat{J}_2(3,2) A_5(s,s_{1234},s'_{13},s'_{134}) + 2\hat{J}_1(3) A_5(s,s_{1234},s'_{14},s'_{124}) + 2\hat{J}_1(3) A_7(s,s_{1234},s'_{24},s'_{124}) +, \quad (8)$$
$$+ 4\hat{J}_1(2) A_6(s,s_{1234},s'_{34},s'_{345})$$

$$A_3(s,s_{1234},s_{25},s_{34}) = \frac{\lambda_3 B_1(s_{25}) B_1(s_{34})}{[1-B_1(s_{25})][1-B_1(s_{34})]} + 6\hat{J}_2(1,1) A_6(s,s_{1234},s'_{23},s'_{234}) +$$
$$+ 6\hat{J}_2(1,1) A_7(s,s_{1234},s'_{24},s'_{234}) + 8\hat{J}_1(1) A_5(s,s_{1234},s'_{13},s'_{134}) \quad (9)$$

$$A_4(s,s_{1234},s_{24},s_{35}) = \frac{\lambda_4 B_2(s_{24}) B_2(s_{35})}{[1-B_2(s_{24})][1-B_2(s_{35})]} + 12\hat{J}_2(2,2) A_6(s,s_{1234},s'_{23},s'_{234}) +$$
$$+ 8\hat{J}_1(2) A_5(s,s_{1234},s'_{13},s'_{135}) \quad (10)$$

$$A_5(s,s_{1234},s_{13},s_{134}) = \frac{\lambda_5 B_3(s_{13})}{1-B_3(s_{13})} + 4\hat{J}_3(3) A_2(s,s_{1234},s'_{12},s'_{35}) + 8\hat{J}_3(3) A_1(s,s_{1234},s'_{12},s'_{34}), \quad (11)$$

$$A_6(s,s_{1234},s_{23},s_{234}) = \frac{\lambda_6 B_1(s_{23})}{1-B_1(s_{23})} + 2\hat{J}_3(1) A_3(s,s_{1234},s'_{34},s'_{25}) + 2\hat{J}_3(1) A_4(s,s_{1234},s'_{24},s'_{35}) +$$
$$+ 2\hat{J}_3(1) A_1(s,s_{1234},s'_{12},s'_{34}) + 2\hat{J}_3(1) A_2(s,s_{1234},s'_{13},s'_{24}) \quad (12)$$

$$A_7(s,s_{1234},s_{24},s_{234}) = \frac{\lambda_7 B_2(s_{24})}{1-B_2(s_{24})} + 4\hat{J}_3(2) A_3(s,s_{1234},s'_{34},s'_{25}) + 4\hat{J}_3(2) A_1(s,s_{1234},s'_{12},s'_{34}), \quad (13)$$

were $\lambda_i$ are the current constants. They do not affect the mass spectrum of pentaquarks. We introduce the integral operators:

$$\hat{J}_1(l) = \frac{G_l(s_{12})}{[1-B_l(s_{12})]} \int_{(m_1+m_2)^2}^{\Lambda_l} \frac{ds'_{12}}{\pi} \frac{G_l(s'_{12})\rho_l(s'_{12})}{s'_{12}-s_{12}} \int_{-1}^{+1} \frac{dz_1}{2}, \quad (14)$$

$$\hat{J}_2(l,p) = \frac{G_l(s_{12}) G_p(s_{34})}{[1-B_l(s_{12})][1-B_p(s_{34})]} \times$$
$$\times \int_{(m_1+m_2)^2}^{\Lambda_l} \frac{ds'_{12}}{\pi} \frac{G_l(s'_{12})\rho_l(s'_{12})}{s'_{12}-s_{12}} \int_{(m_3+m_4)^2}^{\Lambda_p} \frac{ds'_{34}}{\pi} \frac{G_p(s'_{34})\rho_p(s'_{34})}{s'_{34}-s_{34}} \int_{-1}^{+1} \frac{dz_3}{2} \int_{-1}^{+1} \frac{dz_4}{2}, \quad (15)$$

$$\hat{J}_3(l) = \frac{G_l(s_{12},\tilde{\Lambda})}{1-B_l(s_{12},\tilde{\Lambda})} \times$$
$$\times \frac{1}{4\pi} \int_{(m_1+m_2)^2}^{\tilde{\Lambda}} \frac{ds'_{12}}{\pi} \frac{G_l(s'_{12},\tilde{\Lambda})\rho_l(s'_{12})}{s'_{12}-s_{12}} \int_{-1}^{+1} \frac{dz_1}{2} \int_{-1}^{+1} dz \int_{z_2^-}^{z_2^+} dz_2 \frac{1}{\sqrt{1-z^2-z_1^2-z_2^2+2zz_1z_2}}, \quad (16)$$

here $l, p$ are equal 1 - 3. Here $m_i$ is a quark mass.

In the Eqs. (14) and (16) $z_1$ is the cosine of the angle between the relative momentum of the particles 1 and 2 in the intermediate state and the momentum of the particle 3 in the final state, taken in the c.m. of particles 1 and 2. In the Eq. (16) $z$ is the cosine of the angle between the momenta of the particles 3 and 4 in the final state, taken in the c.m. of particles 1 and 2. $z_2$ is the cosine of the angle between the relative momentum of particles 1 and 2 in the intermediate state and the momentum of the particle 4 in the final state, is taken in the c.m. of particles 1 and 2. In the Eq. (15): $z_3$ is the cosine of the angle between relative momentum of particles 1 and 2 in the intermediate state and the relative momentum of particles 3 and 4 in the intermediate state, taken in the c.m. of particles 1 and 2. $z_4$ is the cosine of the angle between the relative momentum of the particles 3 and 4 in the intermediate state and that of the momentum of the particle 1 in the intermediate state, taken in the c.m. of particles 3, 4.

We can pass from the integration over the cosines of the angles to the integration over the subenergies.

Let us extract two-particle singularities in the amplitudes $A_1(s, s_{1234}, s_{12}, s_{34})$, $A_2(s, s_{1234}, s_{12}, s_{35})$, $A_3(s, s_{1234}, s_{25}, s_{34})$, $A_4(s, s_{1234}, s_{24}, s_{35})$, $A_5(s, s_{1234}, s_{13}, s_{134})$, $A_6(s, s_{1234}, s_{23}, s_{234})$, $A_7(s, s_{1234}, s_{24}, s_{234})$:

$$A_1(s, s_{1234}, s_{12}, s_{34}) = \frac{\alpha_1(s, s_{1234}, s_{12}, s_{34}) B_3(s_{12}) B_1(s_{34})}{[1 - B_3(s_{12})][1 - B_1(s_{34})]}, \quad (17)$$

$$A_2(s, s_{1234}, s_{12}, s_{35}) = \frac{\alpha_2(s, s_{1234}, s_{12}, s_{35}) B_3(s_{12}) B_2(s_{35})}{[1 - B_3(s_{12})][1 - B_2(s_{35})]}, \quad (18)$$

$$A_3(s, s_{1234}, s_{25}, s_{34}) = \frac{\alpha_3(s, s_{1234}, s_{25}, s_{34}) B_1(s_{25}) B_1(s_{34})}{[1 - B_1(s_{25})][1 - B_1(s_{34})]}, \quad (19)$$

$$A_4(s, s_{1234}, s_{24}, s_{35}) = \frac{\alpha_4(s, s_{1234}, s_{24}, s_{35}) B_2(s_{24}) B_2(s_{35})}{[1 - B_2(s_{24})][1 - B_2(s_{35})]}, \quad (20)$$

$$A_5(s, s_{1234}, s_{13}, s_{134}) = \frac{\alpha_5(s, s_{1234}, s_{13}, s_{134}) B_3(s_{13})}{1 - B_3(s_{13})}, \quad (21)$$

$$A_6(s, s_{1234}, s_{23}, s_{234}) = \frac{\alpha_6(s, s_{1234}, s_{23}, s_{234}) B_1(s_{23})}{1 - B_1(s_{23})}, \quad (22)$$

$$A_7(s, s_{1234}, s_{24}, s_{234}) = \frac{\alpha_7(s, s_{1234}, s_{24}, s_{234}) B_2(s_{24})}{1 - B_2(s_{24})}, \tag{23}$$

We do not extract three- and four-particle singularities, because they are weaker than two-particle singularities.

We used the classification of singularities, which was proposed in paper [49]. The construction of approximate solution of Eqs. (7) - (13) is based on the extraction of the leading singularities of the amplitudes. The main singularities in $s_{ik} \approx (m_i + m_k)^2$ are from pair rescattering of the particles i and k. First of all there are threshold square-root singularities. Also possible are pole singularities which correspond to the bound states. The diagrams of Fig. 1 apart from two-particle singularities have the triangular singularities, the singularities defining the interaction of four and five particles. Such classification allows us to search the corresponding solution of Eqs. (7) - (13) by taking into account some definite number of leading singularities and neglecting all the weaker ones. We consider the approximation which defines two-particle, triangle, four- and five-particle singularities. The functions $\alpha_1(s, s_{1234}, s_{12}, s_{34})$, $\alpha_2(s, s_{1234}, s_{12}, s_{35})$, $\alpha_3(s, s_{1234}, s_{25}, s_{34})$, $\alpha_4(s, s_{1234}, s_{24}, s_{35})$, $\alpha_5(s, s_{1234}, s_{13}, s_{134})$, $\alpha_6(s, s_{1234}, s_{23}, s_{234})$, $\alpha_7(s, s_{1234}, s_{24}, s_{234})$ are smooth functions of $s_{ik}$, $s_{ijk}$, $s_{ijkl}$, $s$ as compared with the singular part of the amplitudes, hence they can be expanded in a series in the singularity point and only the first term of this series should be employed further. Using this classification one define the reduced amplitudes $\alpha_1$, $\alpha_2$, $\alpha_3$, $\alpha_4$, $\alpha_5$, $\alpha_6$, $\alpha_7$ as well as the B-functions in the middle point of the physical region of Dalitz-plot at the point $s_0$:

$$s_0^{ik} = s_0 = \frac{s + 3\sum_{i=1}^{5} m_i^2}{0.25 \sum_{\substack{i,k=1 \\ i \neq k}}^{5} (m_i + m_k)^2} \tag{24}$$

$$s_{123} = 0.25 s_0 \sum_{\substack{i,k=1 \\ i \neq k}}^{3} (m_i + m_k)^2 - \sum_{i=1}^{3} m_i^2, \quad s_{1234} = 0.25 s_0 \sum_{\substack{i,k=1 \\ i \neq k}}^{4} (m_i + m_k)^2 - 2\sum_{i=1}^{4} m_i^2$$

The approximation (Eq. 24) take into account the definition of two-particle subenergy squared $s_{ik} = (p_i + p_k)^2$, the three-particle energy squared $s_{ijk} = (p_i + p_j + p_k)^2$, the four-particle subenergy squared $s_{ijkl} = (p_i + p_j + p_k + p_l)^2$ and the system total energy squared $s = (p_i + p_j + p_k + p_l + p_m)^2$, here $p_i, p_j, p_k, p_l, p_m$ are the particle momenta. We shall assume that the dimensionless parameter $s_0$ is constant and independent of the type of interacting quarks.

Such a choice of point $s_0$ allows one to replace the integral Eqs. (7) - (13) (Fig. 1) by the algebraic equations (25) - (31) respectively:

$$\alpha_1 = \lambda_1 + 3\alpha_6 J_2(3,1,1) + 3\alpha_7 J_2(3,1,2) + 2\alpha_5 J_2(3,1,3) + 2\alpha_5 J_1(3,3) + 2\alpha_6 J_1(3,1) + \\ + 2\alpha_7 J_1(1,2) + 2\alpha_6 J_1(1,1)  \quad (25)$$

$$\alpha_2 = \lambda_2 + 6\alpha_6 J_2(3,2,1) + 2\alpha_5 J_2(3,2,3) + 2\alpha_5 J_1(3,3) + 2\alpha_7 J_1(3,2) + 4\alpha_6 J_1(2,1), \quad (26)$$

$$\alpha_3 = \lambda_3 + 6\alpha_6 J_2(1,1,1) + 6\alpha_7 J_2(1,1,2) + 8\alpha_5 J_1(1,3), \quad (27)$$

$$\alpha_4 = \lambda_4 + 12\alpha_6 J_2(2,2,1) + 8\alpha_5 J_1(2,3), \quad (28)$$

$$\alpha_5 = \lambda_5 + 4\alpha_2 J_3(3,3,2) + 8\alpha_1 J_3(3,3,1), \quad (29)$$

$$\alpha_6 = \lambda_6 + 2\alpha_3 J_3(1,1,1) + 2\alpha_4 J_3(1,2,2) + 2\alpha_1 J_3(1,1,3) + 2\alpha_2 J_3(1,2,3), \quad (30)$$

$$\alpha_7 = \lambda_7 + 4\alpha_3 J_3(2,1,1) + 4\alpha_1 J_3(2,1,3), \quad (31)$$

We use the functions $J_1(l, p)$, $J_2(l, p, r)$, $J_3(l, p, r)$ ($l, p, r = 1, 2, 3$):

$$J_1(l,p) = \frac{G_l^2(s_0^{12}) B_p(s_0^{13})}{B_l(s_0^{12})} \int_{(m_1+m_2)^2}^{\Lambda_l} \frac{ds'_{12}}{\pi} \frac{\rho_l(s'_{12})}{s'_{12} - s_0^{12}} \int_{-1}^{+1} \frac{dz_1}{2} \frac{1}{1 - B_p(s'_{13})}, \quad (32)$$

$$J_2(l,p,r) = \frac{G_l^2(s_0^{12}) G_p^2(s_0^{34}) B_r(s_0^{13})}{B_l(s_0^{12}) B_p(s_0^{34})} \times \\ \times \int_{(m_1+m_2)^2}^{\Lambda_l} \frac{ds'_{12}}{\pi} \frac{\rho_l(s'_{12})}{s'_{12} - s_0^{12}} \int_{(m_3+m_4)^2}^{\Lambda_p} \frac{ds'_{34}}{\pi} \frac{\rho_p(s'_{34})}{s'_{34} - s_0^{34}} \int_{-1}^{+1} \frac{dz_3}{2} \int_{-1}^{+1} \frac{dz_4}{2} \frac{1}{1 - B_r(s'_{13})} \quad (33)$$

$$J_3(l,p,r) = \frac{G_l^2(s_0^{12},\tilde{\Lambda}) B_p(s_0^{13}) B_r(s_0^{24})}{1 - B_l(s_0^{12},\tilde{\Lambda})} \frac{1 - B_l(s_0^{12})}{B_l(s_0^{12})} \times $$
$$\times \frac{1}{4\pi} \int_{(m_1+m_2)^2}^{\tilde{\Lambda}} \frac{ds'_{12}}{\pi} \frac{\rho_l(s'_{12})}{s'_{12} - s_0^{12}} \int_{-1}^{+1} \frac{dz_1}{2} \int_{-1}^{+1} dz \int_{z_2^-}^{z_2^+} dz_2 \frac{1}{\sqrt{1 - z^2 - z_1^2 - z_2^2 + 2zz_1z_2}} \frac{1}{[1 - B_p(s'_{13})][1 - B_r(s'_{24})]} \tag{34}$$

The other choices of point $s_0$ do not change essentially the contributions of $\alpha_1$, $\alpha_2$, $\alpha_3$, $\alpha_4$, $\alpha_5$, $\alpha_6$, $\alpha_7$, therefore we omit the indexes $s_0^{ik}$. Since the vertex functions depend only slightly on energy it is possible to treat them as constants in our approximation.

The solutions of the system of equations are considered as:

$$\alpha_i(s) = F_i(s, \lambda_i) / D(s), \tag{35}$$

where zeros of $D(s)$ determinants define the masses of bound states of pentaquark baryons. $F_i(s, \lambda_i)$ are the functions of $s$ and $\lambda_i$. The functions $F_i(s, \lambda_i)$ determine the contributions of subamplitudes to the pentaquark baryon amplitude.

### III. Calculation results.

The poles of the reduced amplitudes $\alpha_1$, $\alpha_2$, $\alpha_3$, $\alpha_4$, $\alpha_5$, $\alpha_6$, $\alpha_7$ correspond to the bound states and determine the masses of $\theta^+$ pentaquarks (Fig. 1). If we consider the $\theta^{++}$ ($\theta^0$) (Fig. 2), we also must take into account the interaction of the quarks in the $0^+$ and $1^+$ states. The quark masses of model $m_{u,d} = 410$ MeV and $m_s = 557$ MeV coincide with the quark masses of the ordinary baryons in our model [36]. The model in consideration has not new parameters as compared to previous paper [35]. The dimensionless parameter $g = 0.456$ is determined by fixing of $\theta^+$ pentaquark mass (1540 MeV). The cut-off parameters coincide with those in paper [37]: $\Lambda_{0^+} = 16.5$ and $\Lambda_{1^+} = 20.12$ for the diquarks with $0^+$ and $1^+$ respectively. The cut-off parameter for the mesons is equal to $\Lambda = 16.5$. The masses of $\theta^+$ and $\theta^{++}$ pentaquarks with the quantum number I = 0, 1 and $J^P = \frac{3}{2}^-, \frac{5}{2}^-$ are predicted (Table I). The mass of the lowest pentaquark with I = 0 and $J^P = \frac{3}{2}^-$ is calculated. It is equal to 1514 MeV.

We predict the degeneracy $\theta^{++}$ states with I = 1 and $J^P = \frac{3}{2}^-, \frac{5}{2}^-$. We calculated the contributions of the subamplitudes to the amplitude $\theta^+$ (Table II).

## IV. Conclusion.

In strongly bound system of light quarks such as the baryons where $p/m \sim 1$, the approximation of non-relativistic kinematics and dynamics is not justified. In our relativistic five-quark model (Faddeev-Yakubovsky - type approach) the masses of $\theta$ pentaquarks are calculated. The mass of lowest negative parity pentaquark is equal to 1514 MeV. Capstick, Page and Roberts [17] proposed the isotensor resonance with $J^P = \frac{1}{2}^-, \frac{3}{2}^-, \frac{5}{2}^-$, decaying to K+n. We consider the lowest pentaquark with I = 0 and $J^P = \frac{3}{2}^-$, decaying to $K_S^0 p$.

But the similar result can be obtained for the decay to $K^+ n$ if we take into account the isotopic symmetry. The $\theta^+$ strong overlap with $K_S^0 p$ is equal to 0.17. If $\theta^+$ carries $J^P = \frac{3}{2}^-$, it decays into nucleon and kaon in the D-wave. The phase space suppression factor is about $10^{-5}$, therefore the decay width could be well below ten MeV. [ZEUS Collaboration] [50] reported about new $\theta^+$ pentaquark in the $K_S^0 p$ channel with mass $1520 \pm 4$ MeV. This result is very close to our calculations.

The interesting research is the consideration of the isotensor $qqqq\overline{Q}$ states with $q$ an up or down quark and $Q$ a heavy quark ($Q = c, b$). Their decay to $N(q\overline{Q})$ also violates isospin conservation. It is similar to the case of $\theta$ pentaquarks.

Acknowledgments.

One of authors (S.M.G.) is indebted to University of Liege , where a part of this work was complected, for the hospitality. The authors would like to thank T. Barnes, D.I. Diakonov, A. Hosaka, M.V. Polyakov, Fl. Stancu for useful discussions. This research was supported in part by the Russian Ministry of Education, Program "Universities of Russia" under Contract № 01.20.00.06448.

Figure captions.

Fig. 1. Graphic representation of the equations for the five-quark subamplitudes $A_k$ ($k$ =1-7) corresponding to the $\theta^+$ pentaquark.

Fig. 2. Graphic representation of the equations for the five-quark subamplitudes $A_k$ ($k$ =1-6) corresponding to the $\theta^{++}$ ($\theta^0$) pentaquark.

Table I. Low-lying pentaquark masses (MeV).

| $J^P$ | $\theta^+$ ($udud\bar{s}$), Mass, MeV | $\theta^{++}$ ($uuud\bar{s}$), Mass, MeV |
|---|---|---|
| $\frac{3}{2}^-$ | 1514 | 1550 |
| $\frac{5}{2}^-$ | 1600 | 1550 |

Parameters of model: quark mass $m_{u,d}$ = 410 MeV, $m_s$ = 557 MeV;

cut-off parameter $\Lambda_{0^+}$ =16.5, $\Lambda_{1^+}$ =20.12; gluon coupling constant $g$ =0.456.

Table II. The contributions of subamplitudes to the $\theta^+$ ($J^P = \frac{3}{2}^-$) amplitude in %.

| |
|---|
| $M(2^-)qD(0^+)$ =12.08 |
| $M(2^-)qD(1^+)$ =11.91 |
| $D(0^+)\bar{q}D(0^+)$ =7.66 |
| $D(1^+)\bar{q}D(1^+)$ =30.89 |
| $qqqq\bar{q}$ =37.46 |

Table III. Vertex functions

| $J^{PC}$ | $G_n^2$ |
|---|---|
| $0^+$ (n=1) | $4g/3 - 2g(m_i + m_k)^2/(3s_{ik})$ |
| $1^+$ (n=2) | $2g/3$ |
| $2^-$ (n=3) | $4g/3$ |

Table IV. Coefficient of Chew-Mandelstam functions for n = 3 (meson state) and diquarks n = 1 ($J^P = 0^+$), n = 2 ($J^P = 1^+$).

| $J^{PC}$ | n | $\alpha(J^{PC},n)$ | $\beta(J^{PC},n)$ |
|---|---|---|---|
| $2^-$ | 3 | 4/7 | -1/14-3e/7 |
| $0^+$ | 1 | 1/2 | -e/2 |
| $1^+$ | 2 | 1/3 | 1/6-e/3 |

$e = (m_i - m_k)^2 / (m_i + m_k)^2$

References.

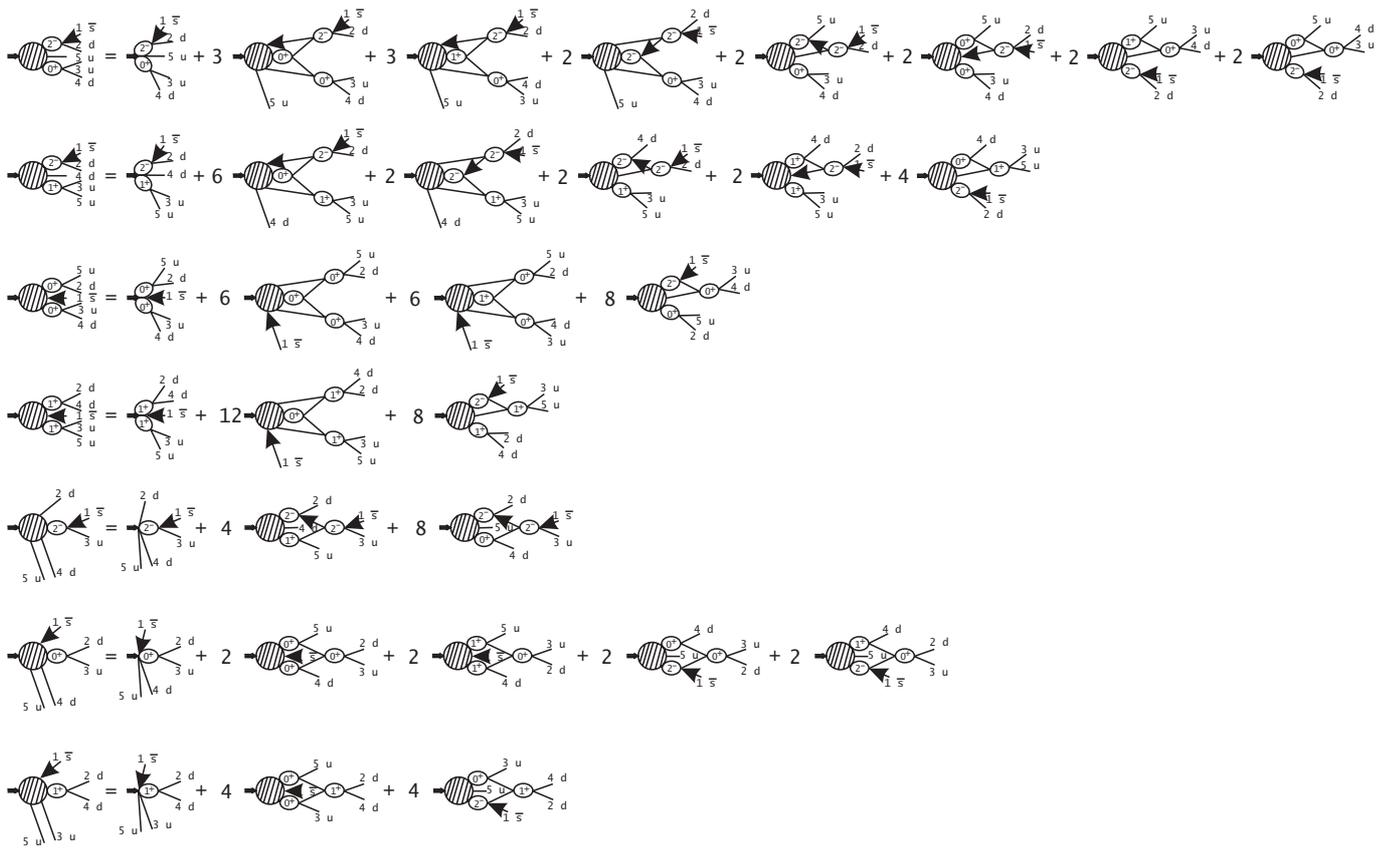

Fig.1

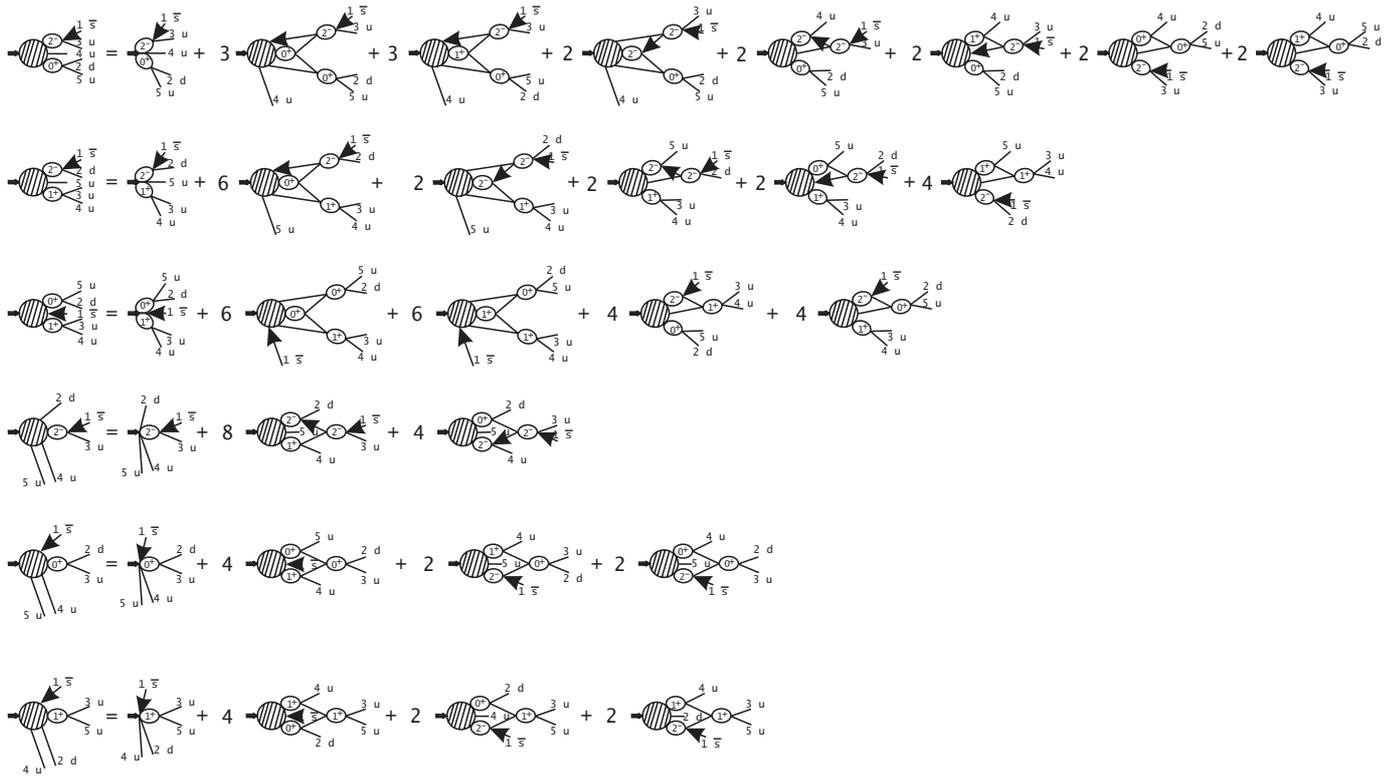

Fig.2